\documentstyle[12pt]{article}
\newcommand{\vect}[1]{\mbox{\bf #1}}

\newcommand{\rms}[1]{\mbox{\scriptsize #1}}

\newcommand{\lw}[1]{\smash{\lower 1.5ex\hbox{#1}}}

\begin{document}
%
%
\title{\Large\bf
Final state interactions for Bose-Einstein correlations in S+Pb $\to$
$\pi^+$$\pi^+$ +X reaction \\at energy 200GeV/nucleon \vspace{-2mm}\\}
\author{T. Osada$^1$
\thanks{e-mail: osada@nucl.phys.tohoku.ac.jp}~
\thanks{permanent address: Department of Physics, 
Tohoku University, Sendai 980, Japan}~,
~S. Sano$^1$,~ M. Biyajima$^1$
\thanks{e-mail: minoru44@jpnyitp.bitnet}~
and G.Wilk$^2$ \thanks{e-mail: wilk@fuw.edu.pl}\\
{\small $^1$ Department of Physics, Faculty of Science,
Shinshu University, Matsumoto 390, Japan}\\
{\small $^2$ Soltan Institute for Nuclear Studies, Zd-PVIII,
Ho\.za 69, PL-00-681 Warsaw, Poland }\\
}
\date{\today}
\maketitle
\begin{abstract}
We applied an analytical formula for Bose-Einstein correlations (BEC)
developed by us recently to high-energy heavy ion collisions, in
particular to data on S+Pb$\to\pi^+$-$\pi^+$+X reaction at energy
$200$ GeV/nucleon reported by the NA44 Collaboration. It takes into
account both Coulomb and strong ($\pi$-$\pi$~s-wave; I=2) final state
interactions (FSI). We have found that inclusion of the strong interaction 
in addition to Coulomb correction affects significantly the extracted
parameters of the BEC like the source size $R$, the degree of coherence
$\lambda$ and the long-range correlation parameter $\gamma$. In
particular, the $\lambda$ parameter of the BEC is increased by about
20\%. Our results differ from those obtained in $e^+e^-$ annihilation
in the following way: the $\lambda$ parameter does not reach `chaotic
limit' and the $\gamma$ parameter does not approach to zero.  
\end{abstract}
\newpage
\noindent 
{\bf Introduction:}~~~
Recently we have presented an analytical formula for Bose-Einstein
correlations (BEC), which takes into account both Coulomb and strong
($\pi$-$\pi$~s-wave; I=2) final state interactions (FSI) and applied
it to $e^+e^-$ annihilation data~\cite{ZPHYS96}. Our formula has been
constructed by using a wave function obtained from applying the phase 
shift of the strong interaction to the asymptotic Coulomb wave function, 
and extrapolating it to the non-asymptotic region with the normalization 
of the Coulomb wave function. 
We shall apply it here to analysis of high energy nuclear collisions 
using in addition the following important parameters entering the BEC: 
degree of coherence $\lambda$, long range correlation
$\gamma$ and normalization $c$. This will make our analysis of the FSI
much more clear when compared with a `standard' fitting formula with or 
without the Coulomb corrections. 
In the following sections we first provide brief explanation of our 
formula, use it next to investigate data for S+Pb$\to\pi^+$-$\pi^+$~+X 
collisions at energy $200$GeV/nucleon reported by the 
NA44 Collaboration\cite{NA44-SPb} and, finally, we close our discussion 
with some concluding remarks.\\ 

\noindent 
{\bf Coulomb and strong wave function with s-wave phase shift:}~~~
Let us consider the Coulomb asymptotic wave function for the identical
$\pi$-$\pi$ scattering (of relative momentum $Q$, the pion mass is
$m_{\pi}$). Its form is well known and given by
\cite{SCHIFF,SASAKAWA}: 
\begin{eqnarray}
\Psi_{\rms{C}}^{\rms{asym}} (\vect{k}, \vect{r})
&=& \exp \{ i(kz + \eta\ln(k(r-z))) \}
\left[ 1 + \frac{\eta^2}{ik(r-z)} \right] \nonumber \\
& & + f(\theta) \frac{\exp\{i(kr - \eta \ln(2kr))\}}r\:,
\end{eqnarray}
\noindent 
where $Q=2k,~z = r \cos \theta$, $\eta=m \alpha/Q$ and the scattering
amplitude~$f(\theta)$~is given by:
\begin{eqnarray*}
f(\theta) = - \frac{\eta}{2k} \frac 1{\sin^2 (\theta/2)}
\exp\{ -2i\eta\ln \sin (\theta/2) + 2i \arg
\Gamma (1+i\eta) \}.
\end{eqnarray*}
\noindent 
The s-wave component of the $\Psi_{\rms{C}}^{\rm{asym}}(\vect{k},\vect{r})$ 
wave function is given by the following formula~\cite{SCHIFF,SASAKAWA}:
\begin{eqnarray}
\Psi_{\rms{C(s-wave)}}^{\rms{asym}} (\vect{k}, \vect{r})
&=&   \exp \{ i(kr - \eta\ln(2kr) + 2\eta_0) \} \frac{1}{2ikr}
\left[ 1 + \frac{i\eta(1+i\eta)}{2ikr} \right] \nonumber \\
& & + \exp \{-i(kr - \eta\ln(2kr)) \} \frac{1}{-2ikr}
\left[ 1 + \frac{i\eta(1-i\eta)}{2ikr} \right].
\label{CsAsy}
\end{eqnarray}

The strong wave fucion of the s-wave has been obtained by extracting 
the out-going wave (up to the order ${\cal O}(1/r)$) 
of the eq.(\ref{CsAsy}) and using the phase
shift of the $\pi$-$\pi$ scattering induced by the strong interaction
\cite{WALKER}-\cite{HOOGLAND}. This phase shift is phenomenologically
given by \cite{M-SUZUKI}: 
\begin{eqnarray}
\delta_{0}^{(2)}=\frac{1}{2} \left( \frac{a_0Q}{1+0.5Q^2} \right)
~~\mbox{($ -1.5 \le a_0 \le -0.7 $~(GeV${}^{-1})$~)} .
\end{eqnarray}
As a result we obtain the following strong interaction wave 
function in the asymptotic region:
\begin{eqnarray}
\phi_{\rms{st}}(\vect{k}, \vect{r})=f^0(\theta) 
\frac{\exp\{i(kr - \eta \ln(2kr))\}}r,\\
f^0(\theta)=\frac{1}{2ik} 
\exp(2i\eta_0) (\exp(2i\delta_{0}^{(2)})-1).\nonumber
\end{eqnarray}
Therefore, the total asymptotic wave function is expressed as the
following sum of both components:
\begin{eqnarray}
\Psi_{\rms{total}}(\vect{k},\vect{r}) =
\Psi_{\rms{C}}^{\rms{asym}}(\vect{k},\vect{r})
+ \phi_{\rms{st}}(\vect{k},\vect{r}). \label{ASYMP}
\end{eqnarray}

In the small $kr$ (internal) region the Coulomb wave function is
given by~\cite{SCHIFF,SASAKAWA} 
\begin{eqnarray}
\Psi_{\rms{C}}(\vect{k},\vect{r})
&=& \Gamma(1+i\eta) e^{-\pi\eta/2}e^{i{\vect{\small{k}}\cdot
\vect{\small{r}}}}F(-i\eta;1;ikr(1-\cos\theta)), 
\end{eqnarray} 
(where $F$ denotes the confluent hypergeometric function) whereas the
exact form of the strong wave function in this region is unknown.
We assume therefore that the strong wave function is given by the
extrapolation of the asymptotic wave function into this internal
region~\cite{MG-BOWLER} with the normalization of the Coulomb wave
function~\cite{ZPHYS96}: 
\begin{eqnarray}
\Psi_{\rms{total}}(\vect{k},\vect{r}) =
\Psi_{\rms{C}}(\vect{k},\vect{r})
+ \sqrt{G(\eta)}\phi_{\rms{st}}(\vect{k},\vect{r}), \label{INTER}
\end{eqnarray}
where $G(\eta)$ denotes the usual Gamow factor:
$G(\eta)=2\pi\eta/(\exp(2\pi\eta)-1)$ (for the discussion on the
validity and usefulness of this assumption and on the smooth
connection of wave functions given by eqs.(\ref{ASYMP}) and
(\ref{INTER}) cf. Ref.~\cite{ZPHYS96}). Because in the numerical
calculations of the $\Psi_{\rms{C}}(\vect{k},\vect{r})$ one encounters a
wild oscillating behavior in larger $kr$ region, we use here the
`seamless' fitting method developed in \cite{SEAMLESS}.\\

\noindent 
{\bf Theoretical formula of BEC with Coulomb and strong wave function:}~~
To describe a pair of the identical bosons, we have to symmetrize its
total wave function in the following way:  
\begin{eqnarray}
A_{12} = \frac{1}{\sqrt{2}}[ \Psi_{\rms{C}}(\vect{k},\vect{r})
+ \Psi_{\rms{C}}^{S}(\vect{k},\vect{r})+
\Phi_{\rms{st}}(\vect{k},\vect{r})+
\Phi_{\rms{st}}^{S}(\vect{k},\vect{r})],
\end{eqnarray}
where the superscript $S$ denotes the symmetrization of the corresponding
wave function. Here $ \Phi_{\rms{st}}(\vect{k},\vect{r}) $~stands for
the wave function induced by the strong interactions. Assuming that
a source function is given by $\rho(r)$ we obtain the following 
expression for the BEC including the FSI~\cite{ZPHYS96}: 
\begin{eqnarray}
  N^{(\pm\pm)}/N^{BG} &=& 
   \frac{1}{G(\eta)} \int_{}^{}
   \rho(r) d^3r|A_{12}|^2
\equiv I_{\rms{C}} + I_{\rms{Cst}} + I_{\rms{st}}, \label{STCL-formula}\\
&&\!\!\!\!\!\!\!\!\!\!\!\!\!\!\!\!\!\!\!\!\!\!\!\!
\!\!\!\!\!\!\!\!\!\!\!\!\!\!\!\!\!\!\!\!\!\!\!\!\!
 I_{{\rms C}}  = \sum_{m,n=0}^{\infty}
               \frac{1}{m+n+1}
       I_{R1}(2+m+n) A_1(n)A_1^{\ast}(m)
           \times \left[
                      1+\frac{n!m!}{(n+m)!}
                     \left( 1+\frac{n}{i\eta} \right)
                     \left( 1-\frac{m}{i\eta} \right)
               \right] \nonumber \\
&&\!\!\!\!\!\!\!\!\!\!\!\!\!\!\!\!\!\!\!\!\!\!\!\!
\!\!\!\!\!\!\!\!\!\!\!\!\!\!\!\!\!
=(1 + \Delta_{\rms{1C}}) + (E_{2\rms{B}} +
\Delta_{\rms{EC}}), \nonumber \\
&&\!\!\!\!\!\!\!\!\!\!\!\!\!\!\!\!\!\!\!\!\!\!\!\!\!\!
\!\!\!\!\!\!\!\!\!\!\!\!\!\!\!\!\!\!\!\!\!\!\!\!\!\!
I_{{\rms Cst}}=2\Re
               \left[
        \frac{2}{k} (2k)^{i\eta}
   \exp{(-i(\eta_0+\delta_{0}^{(2)}))} \sin\delta_{0}^{(2)}
    \sum_{n=0}^{\infty} I_{\rms{R2}}(1+n) A_2(n,0)
               \right], \nonumber \\
&&\!\!\!\!\!\!\!\!\!\!\!\!\!\!\!\!\!\!\!\!\!\!\!\!
\!\!\!\!\!\!\!\!\!\!\!\!\!\!\!\!\!\!\!\!\!\!\!\!\!
I_{{\rms st}}=\frac{2}{k^2}
I_{\rms{R1}}(0) \sin^2\delta_{0}^{(2)}\nonumber,
\end{eqnarray}
where
\begin{eqnarray*}
   E_{2{\rms B}} &=& \int_{}^{} d^3r~
\rho(r) ~
   {\textstyle e}^{-{\rms i}
   {\scriptstyle {\bf Q}} \cdot
   {\scriptstyle {\bf r}} }, \\
   1 + \Delta_{1{\rms C}} &=& 1+4\pi \cdot 2\eta
                          \int_{}^{}\rho(r) r^2dr
     \sum_{n=0}^{\infty}\frac{(-1)^n A^{2n+1}}{(2n+1)!
(2n+1)(2n+2)}, \\
   A_1(n)    &=& \frac{\Gamma(n+i\eta)}{\Gamma(i\eta)}
                 \frac{(-2ik)^n}{(n!)^2}, \\
   A_2(n,l)  &=& \frac{\Gamma(n+l+1+i\eta)\Gamma(2l+2)}
                      {\Gamma(l+1+i\eta)\Gamma(n+2l+2)}
                 \frac{(-2ik)^n}{n!}, \\
   I_{{\rms R1}}(n) &=& 4\pi \int_{}^{}dr r^n \rho(r), \\
   I_{{\rms R2}}(n) &=& 4\pi \int_{}^{}dr r^{n+i\eta} \rho(r).
\end{eqnarray*}
In this paper $N^{(\pm\pm)}/N^{BG}$ 
stands for the FSI corrected ratio of pairs of the identical 
charged bosons measured in a single event to a product of single 
bosons taken from different events. The $\Delta_{\mbox{\footnotesize
1C}}$ and $\Delta_{\mbox{\footnotesize EC}}$  originate from the
correction for the finite size effects in the Coulomb interaction. 
In the following we shall use an artificial Gaussian form of the source
function: $\rho(r)=(\frac{1}{\sqrt{2 \pi } \beta})^3\exp
(\frac{-r^2}{2 \beta^2})$, which results in the following form for
its Fourier transform: 
$$ E_{2 {\rms B}}=\exp(-\beta^2 Q^2/2).$$

\vspace*{10mm}
\noindent 
{\bf Reanalyses of data reported by NA44 Collaboration:}~~~
The parameters mentioned at the beginning, namely: degree of coherence
$\lambda$, long range correlation $\gamma$ and normalization $c$, are
introduced into our formula eq.(\ref{STCL-formula}) in the usual way
leading to the following final formula:
\begin{eqnarray}
N^{(\pm\pm)}/N^{\rms{BG}}
(Q = 2k) &=& c~( 1 + \Delta_{\rms{1C}}
+\Delta_{\rms{EC}}+ I_{\rms{Cst}} + I_{\rms{st}}) \nonumber \\
&&\times\left[1 +\lambda \frac{E_{\rms{2B}}}
{1+\Delta_{\rms{1C}}+\Delta_{\rms{EC}}+I_{\rms{Cst}}+
I_{\rms{st}}}\right](1+\gamma Q)~. \hspace{1cm}\label{STCL-ana}
\end{eqnarray}

This formula have been applied to data on S+Pb $\to \pi^+\pi^+$~+X 
reaction at energy $200$ GeV/nucleon reported by the NA44
Collaboration~\cite{NA44-SPb}.\\

Before presenting our results based on eq.(\ref{STCL-ana}) we
must, however, first clarify one point concerning aplication of
the Coulomb correction. We show the results in Table I obtained by applying 
the following pure Coulomb (i.e., without strong interactions) corrected 
formula:
\begin{eqnarray}
N^{(\pm\pm)}/N^{\rms{BG}}=
c~( 1 + \Delta_{\rms{1C}}+\Delta_{\rms{EC}}) 
\times\left[1 +\lambda \frac{E_{\rms{2B}}}
{1+\Delta_{\rms{1C}}+\Delta_{\rms{EC}}}
\right](1+\gamma Q),\hspace{1cm}\label{CL-ana}
\end{eqnarray}to the BEC data presented by the NA44 Collaboration. 
On the other hand, the NA44 Collaboration 
have also corrected their data
by Coulomb correction factors, $K_{Coul}^{\mbox{\scriptsize NA44}}$,
which have been calculated by using the Coulomb wave function
integration. However, as is shown in the Table I, slight discrepancies
are found between their results (obtained by using the
$K_{Coul}^{\mbox{\scriptsize NA44}}$) and ours \big(obtained from minimum
$\chi^2$ fit using formula eq.(\ref{CL-ana})\big)
\footnote{To compare our estimated values with those reported by the 
the NA44 Collaboration, we have to use a relation, $\beta=\sqrt{2}R$ 
\big( which reflects the fact that NA44 Collaboration employed the following 
fitting formula: $c~(1+\lambda \exp(-Q^2R^2))~$\big).}.
The origin of discrepancy shown in the Table I ~can be traced down to the 
15\% differences between the $K_{Coul}^{\mbox{\scriptsize NA44}}$ and the 
$K_{Coul}^{\mbox{\scriptsize our}}$ (see Table II) (one can see this
by evaluating the correction factor, $K_{Coul}^{\mbox{\scriptsize
our}} =1/G(\eta)[1+\Delta_{\rms{1C}}+\Delta_{\rms{EC}}]$). Our final 
results obtained by using eq.(\ref{STCL-ana}) are shown in Tables III, 
IV and in Fig.1. For the sake of reference the results 
of eq.(\ref{CL-ana}) and of the `standard' fitting formula: 
\begin{eqnarray}
N^{(\pm\pm)}/N^{\rms{BG}}=c\left[1 +\lambda 
E_{\rms{2B}}\right](1+\gamma Q)\label{Standard}, 
\end{eqnarray}
are also shown there. It is found that the FSI corrections, which take 
into account not only the Coulomb but also the strong interactions between
the identical charged pions, affect substantially the source size
parameter $R$, the degree of coherence parameter $\lambda$ and the
long-range correlation parameter $\gamma$. When compared with the
Coulomb correction case only (i.e., with eq.(11)), the $R$ parameter is
reduced by about 5\% and the $\lambda$ parameter is increased by
about 20\% remaining, however, always below its `chaotic' limit value
of $\lambda =1$\cite{ZPHYS96} (notice that in $e^+e^-$ annihilation
the corrections caused by the FSI pushes the $\lambda$ towards this limit).
In what concerns the parameter $\gamma$, we have found that it does not
reduce to zero (differently than in the $e^+e^-$ annihilation case).\\

\noindent
{\bf Concluding remarks:}~~~
We applied our improved formula for the BEC (which accounts for the FSI) 
to data for S+Pb $\to \pi^+\pi^+$~+X~at energy $200$GeV/nucleon
presented recently by the NA44 Collaboration. We have found that 
inclusion of the full FSI \big(including also the strong interactions, cf
eq.(\ref{STCL-ana})\big) increases the extracted value of the degree
of coherence parameter $\lambda$ by about 20\% in comparison to the
case where only the Coulomb correction are applied (cf. eq.(\ref{CL-ana})).
However, differently than in $e^+e^-$ annihilation case (cf.
\cite{ZPHYS96}) it does not reach `chaotic'
limit of $\lambda =1$. We have also found that the long-range
correlation parameter $\gamma$ does not approach to zero which also
seems different from the $e^+e^-$ annihilation case. On the other
hand the source size parameter $R$ is decreased by about 5\% only.  

\vspace*{1cm}
\noindent 
{\bf Acknowlegements:}~~~\\
The authors would like to thank T.Mizoguchi for his kind correspondence. 
One of authors (T.O.) would like to thank Shinshu Univ. 
for their hospitalities. 
This work is partially supported by Japanese Grant-in-Aid for Scientific 
Research from the Ministry of Education, Science Sport and
Culture~(\#06640383).  


\vspace{1cm}
\begin{center}
{\bf Figure Captions.}\\
\end{center}
\begin{itemize}
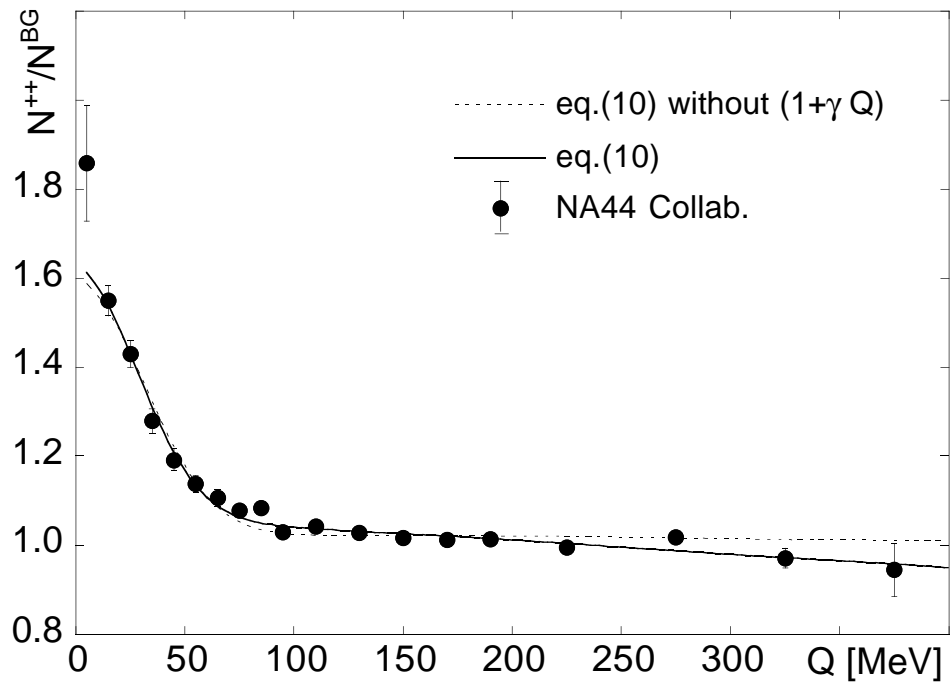
\item[Fig. 1.]
~Analysis of data for S+Pb$\to \pi^+\pi^+$ +X reaction at energy
$200$ GeV/nucleon reported by the NA44 Collaboration\cite{NA44-SPb}
by the minimum $\chi^2$ fitting method. The strong interaction phase
shift parameter has been set to $a_0=-1.20$ GeV$^{-1}$. The solid and dotted
curves represent fitted result obtained by using eq.(\ref{STCL-ana})
with and without factor (1+$\gamma$Q), respectively. 
\end{itemize}

\begin{center}
{\bf Table Captions.}\\
\end{center}
\begin{itemize}
\item[Table 1.]
Results of analysis of BEC data for S+Pb$\to\pi^+\pi^+$ +X 
reaction at energy $200$ GeV/nucleon reported by the NA44 
Collaboration\cite{NA44-SPb}. Comparison of our results obtained from 
fitting of the eq.(11) with values reported by 
the NA44 Collaboration 
which were obtained by using the Coulomb correction factor 
$K_{Coul}^{\mbox{\scriptsize NA44}}$\cite{NA44-SPb}.
\item[Table 2.]
Comparison of the Coulomb correction factor 
$K_{Coul}^{\mbox{\scriptsize NA44}}$ (used in ref.[2]) with \\
the $K_{Coul}^{\mbox{\scriptsize our}}$ (evaluated from eq.(2)).
\item[Table 3.]
Results of analysis of BEC data for S+Pb$\to \pi^+\pi^+$ +X 
reaction at energy $200$ GeV/nucleon reported by the NA44 
Collaboration\cite{NA44-SPb}. 
Comparison of our results obtained from fitting of eqs.(10), (11) 
and (12) without factor $(1+\gamma Q)$. We have employed the parameter 
$a_{0}=-1.20$ GeV$^{-1}$.
\item[Table 4.]
Results of analysis of BEC data for S+Pb$\to \pi^+\pi^+$ +X reaction at 
energy $200$ GeV/nucleon reported by the NA44 Collaboration\cite{NA44-SPb}. 
Comparison of our results obtained from fitting of the 
eqs.(10), (11) and (12). We have employed the parameter 
$a_{0}=-1.20$ GeV$^{-1}$.
\end{itemize}
%
%
\begin{table}[H]
\caption{}
\begin{center}
\begin{tabular}{c|ccccc}
\hline \hline
formula
& c
& $R$ [fm]
& $\lambda$
& $\gamma$
& $\chi^2/$NDF\\
\hline
correction by
&\lw{ 1.000$\pm$0.002}
&\lw{  4.50$\pm$0.431}
&\lw{  0.46$\pm$0.04 }
&\lw{   ---}
&\lw{ 18.1/16}\\
$K_{Coul}^{\mbox{\scriptsize NA44}}$
&&&&&\\
\cline{1-6}
eq.~(11) 
&\lw{  1.000$\pm$0.003}
&\lw{  4.697$\pm$0.272}
&\lw{  0.481$\pm$0.031}
&\lw{  ---}
&\lw{  34.5/16}\\
without $(1+\gamma Q)$
&&&&&\\
\cline{1-6}
\lw{  eq.~(11)}
&\lw{  1.031$\pm$0.009}
&
\lw{  5.159$\pm$0.331}
&\lw{  0.453$\pm$0.034}
&\lw{ $-$0.181$\pm$0.048} 
&\lw{  22.1/15} \\ 
&&&&&\\
\hline
\end{tabular}
\end{center}
\end{table}
%
%
\begin{table}[H]
\caption{}
\begin{center}
\begin{tabular}{ccc}
\hline \hline
Q~[MeV/c]&$K_{Coul}^{\mbox{\scriptsize NA44}}$&
$K_{Coul}^{\mbox{\scriptsize our}}$\\
\hline
5  &1.565&1.807\\
15 &1.161&1.112\\
25 &1.074&1.037\\
35 &1.037&1.014\\
45 &1.020&1.008\\
55 &1.012&1.006\\
65 &1.009&1.006\\
75 &1.007&1.006\\
85 &1.005&1.005\\
95 &1.005&1.004\\
110&1.003&1.003\\
130&1.003&1.002\\
150&1.002&1.000\\
170&1.002&0.999\\
190&1.001&0.999\\
225&1.001&1.000\\
275&1.001&1.003\\
325&1.001&1.004\\
375&1.000&1.007\\
\end{tabular}
\end{center}
\end{table}

%
%
\begin{table}[H]
\caption{}
\begin{center}
\begin{tabular}{l|ccccc}
\hline \hline
formula
& c
& $R$ [fm]
& $\lambda$
& $\gamma$
& $\chi^2/$NDF\\
\hline
eq.~(10)
&\lw{ 1.007$\pm$0.003}
&\lw{ 4.422$\pm$0.282}
&\lw{ 0.593$\pm$0.028}
&\lw{---}
&\lw{ 46.5/16}\\
strong + Coulomb
&&&&&\\
\cline{1-6}
\hline 
eq.~(11)
&\lw{ 1.000$\pm$0.003}
&\lw{ 4.697$\pm$0.272}
&\lw{ 0.481$\pm$0.031}
&\lw{---}
&\lw{ 34.5/16}\\
~~~~~~~Coulomb
&&&&&\\
\cline{1-6}
\hline 
eq.~(12)
&\lw{ 1.019$\pm$0.003}
&\lw{ 4.209$\pm$0.206}
&\lw{ 0.528$\pm$0.031}
&\lw{---}
&\lw{ 48.1/16}\\
~~~~~~~standard
&&&&&\\
\cline{1-6}
\hline 
\end{tabular}
\end{center}
\end{table}
%
%
\begin{table}[H]
\caption{}
\begin{center}
\begin{tabular}{l|ccccc}
\hline \hline
formula
& c
& $R$ [fm]
& $\lambda$
& $\gamma$
& $\chi^2/$NDF\\
\hline
eq.~(10)
&\lw{ 1.045$\pm$0.008}
&\lw{ 4.870$\pm$0.320}
&\lw{ 0.541$\pm$0.031}
&\lw{-0.232$\pm$0.047}
&\lw{ 24.6/15}\\
strong + Coulomb
&&&&&\\
\cline{1-6}
\hline 
eq.~(11)
&\lw{ 1.031$\pm$0.009}
&\lw{ 5.169$\pm$0.331}
&\lw{ 0.453$\pm$0.034}
&\lw{-0.181$\pm$0.048}
&\lw{ 22.1/15}\\
~~~~~~~Coulomb
&&&&&\\
\cline{1-6}
\hline 
eq.~(12)
&\lw{ 1.069$\pm$0.010}
&\lw{ 4.929$\pm$0.292}
&\lw{ 0.514$\pm$0.034}
&\lw{-0.274$\pm$0.051}
&\lw{ 18.1/15}\\
~~~~~~~standard
&&&&&\\
\cline{1-6}
\hline 
\end{tabular}
\end{center}
\end{table}
\end{document}